# Responsible AI in the Software Industry: A Practitioner-Centered Perspective

Matheus de Morais Leça
University of Calgary
Calgary, AB, Canada
matheus.demoraisleca@ucalgary.ca

Mariana Bento
University of Calgary
Calgary, AB, Canada
mariana.pinheirobent@ucalgary.ca

Ronnie de Souza Santos
University of Calgary
Calgary, AB, Canada
ronnie.desouzasantos@ucalgary.ca

*Abstract*—Responsible AI principles provide ethical guidelines for developing AI systems, yet their practical implementation in software engineering lacks thorough investigation. Therefore, this study explores the practices and challenges faced by software practitioners in aligning with these principles. Through semi-structured interviews with 25 practitioners, we investigated their methods, concerns, and strategies for addressing Responsible AI in software development. Our findings reveal that while practitioners frequently address fairness, inclusiveness, and reliability, principles such as transparency and accountability receive comparatively less attention in their practices. This scenario highlights gaps in current strategies and the need for more comprehensive frameworks to fully operationalize Responsible AI principles in software engineering.

## I. INTRODUCTION

Responsible AI is a concept related to the development of artificial intelligence that ensures systems align with ethical values and respect for human rights, emphasizing accountability by the individuals and organizations responsible for creating and deploying these technologies [1]. This concept highlights the accountability of decisions and outcomes generated by algorithms, aiming to build trust, safeguard individual and societal interests, and ensure AI systems contribute positively to human well-being [2], [3].

Responsible AI is grounded in a set of principles that provide ethical guidelines for AI development [4]. *Fairness* ensures that AI systems treat everyone equally and avoid discrimination based on personal characteristics such as race, gender, or religion. *Reliability* and *safety* emphasize that systems must perform consistently and safely across various conditions, fostering trust in their outputs. *Privacy* and *security* are essential for protecting users' confidential information and ensuring systems resist unauthorized access or compromise. *Inclusiveness* addresses barriers to access, empowering all users by incorporating diverse needs into design and fostering innovation. *Transparency* ensures that AI systems are understandable and accessible, allowing individuals to comprehend how decisions are made and their implications. Finally, *accountability* requires developers and organizations to be answerable for their systems' impacts, with mechanisms to monitor, address, and adjust for unintended consequences [5].

Nowadays, with the growing demand for AI-powered systems in various contexts in society, software engineering should incorporate the principles of Responsible AI throughout the software development lifecycle, from planning to testing [6], [7]. For instance, requirements engineering helps identify ethical considerations early, ensuring fairness, privacy, and inclusiveness are included in the system's foundation [7], [8]. Design and testing processes integrate these principles by addressing biases, managing data securely, and promoting transparency in AI operations [9], [10]. Continuous monitoring allows systems to adapt to real-world data and ethical standards, aligning them with societal needs. In this process, software practitioners should use methods and practices that guide AI technologies toward Responsible AI goals [10], [11].

Considering that Responsible AI is an evolving area within software development, in this paper, we explored practitioners' experiences with Responsible AI principles in the context of AI development. Specifically, we examined the practices they are employing to incorporate these principles and the challenges they face in doing so. To guide our research, we formulated the following research questions: ***RQ1.*** *How do software practitioners integrate Responsible AI principles into AI development?* ***RQ2.*** *What challenges do software practitioners encounter in applying Responsible AI principles?*

## II. METHOD

**Industrial Context**. In this study, we interviewed software professionals, including designers, software engineers, data scientists, and testers, from four AI-focused projects at a large South American software company. Founded in 1996, the company serves sectors such as finance, telecommunications, and government, with over 1,200 employees, 70% of whom are directly engaged in software development. The teams are diverse in technical expertise and personal backgrounds, employing methodologies like Scrum, Kanban, and Waterfall to deliver solutions to global clients. Using convenience sampling [12], we selected professionals from four AI projects based on their availability and contractual limitations. These projects developed a range of AI technologies, including deep learning, large language models, and computer vision. While some projects focused on social applications, such as education, others addressed industrial challenges, including those in the energy sector. This context provided a practical setting to investigate how Responsible AI principles are understood and implemented in real-world software development.

**Participants.** We started sampling participants for the study by sending an open invitation to those involved in the selected projects, allowing them to choose convenient interview dates and times, following a convenience sampling approach [12]. Snowball sampling [12] was then employed, with participants recommending colleagues who could contribute to the study. To ensure diverse perspectives, theoretical sampling [13] was used to target individuals with specific expertise, experience levels, or demographic backgrounds. Our study demographics are composed of 25 professionals actively engaged in the development of AI-powered systems, representing a diverse range of roles, including data scientists, software developers, QA/testers, designers, and project managers. Demographically, 36% of participants identified as non-male, including one non-binary software engineer, and 8% had disabilities, such as hearing impairments. Additionally, 16% were non-white, with representation from individuals of African descent, and 24% identified as LGBTQIA+, including gay, lesbian, and queer participants. Furthermore, 16% were neurodivergent, including individuals diagnosed with ADHD. Professionally, 48% held advanced degrees (Master's or Ph.D.), and 44% had over five years of experience in software development. This diverse group provided valuable perspectives on how practitioners navigate and integrate Responsible AI principles in real-world settings.

**Data Collection.** We conducted semi-structured interviews with 25 software practitioners across three rounds between June 1 and July 5, 2024. The first two rounds included seven participants each, with 11 joining in the third. Interviews lasted 23–42 minutes and totaled 4 hours and 49 minutes of audio and 504 pages of transcripts. Three participants who were unable to attend recorded sessions provided responses via open-ended questionnaires. The interviews explored practitioners' routines, methods for identifying and addressing bias, concerns in AI development, skills required, and validation processes to prevent discrimination. Topics included integration challenges, bias identification methods, and fairness management throughout development. Although explicit questions about Responsible AI were not asked, participants frequently addressed its principles indirectly. These connections emerged during qualitative analysis and are detailed in this study.

**Data Analysis.** We used thematic analysis [14] to analyze participants' interviews, identifying patterns and themes within the data. We began with open coding, generating initial codes from transcripts to capture practices and challenges related to implementing Responsible AI principles. To support this phase, we used GPT-4 Omni on the ChatGPT platform, leveraging prompt engineering [15] to extract relevant information. The output was organized in a spreadsheet with participant IDs, aspects mentioned, and corresponding quotes. We verified accuracy by reviewing a random 20% of the extracted data to ensure alignment with the original interviews. Next, we conducted manual axial coding to group related codes into broader categories, focusing on how practitioners apply Responsible AI principles, the challenges they encounter, and their methods. Finally, we used manual selective coding to identify key themes, such as the application of Responsible AI principles, barriers to implementation, and strategies to address these challenges, providing insights into practitioners' experiences in AI development.

**Ethics.** This research followed the ethical guidelines set by the author's university. Participants were informed about the study's purpose, the voluntary nature of their participation, and their right to withdraw at any stage. Informed consent was obtained prior to the interviews, and all data presented in this paper has been anonymized to ensure participant confidentiality.

### III. FINDINGS

Through our findings, Table I and Table II present evidence from the interviews to support our analysis. Some quotations may read oddly as they were directly transcribed and translated from the audio recordings. The code *Pxy* is used to refer to the participants in the sample. More anonymized interview quotations can be accessed via this link: https://figshare.com/s/ddcfa6dcc616d2bd8ea6

#### A. Perception of Responsible AI Principles Among Software Engineering Practitioners

*Fairness* was referenced in participants' narratives as they described concerns about addressing biases in AI systems. Participants characterized *fairness* as the need to ensure equitable treatment of all users, particularly through balanced representation in datasets and avoiding discriminatory outcomes. They referred to *fairness* as an important consideration in mitigating potential harm to users caused by biased systems.

Additionally, we observed that participants referred to *reliability* and *safety* as the ability of AI systems to perform consistently and predictably across different conditions. Participants described these principles as necessary for maintaining trust in AI outputs and emphasized the importance of iterative processes in identifying and addressing performance issues. *Reliability* and *safety* were seen as integral to ensuring that systems meet stakeholder expectations.

*Inclusiveness* was central in participants' narratives through their descriptions of designing AI systems to accommodate diverse users and contexts. Participants referred to *inclusiveness* as a principle that involves considering underrepresented groups in the process, promoting accessibility, and ensuring that AI systems serve the needs of a wide range of users. This principle was often discussed in relation to creating equitable and fair systems.

Participants referred to *transparency* while describing their ability or challenges in understanding and interpreting the outputs and processes of AI systems. They described *transparency* as an important factor in ensuring that stakeholders can audit and trust the decisions made by AI models. *Transparency* was seen as critical for clarifying how systems function and addressing potential concerns about their behavior.

We also observed references to *accountability* in participants' discussions about managing stakeholder relationships and responding to challenges during AI development. Participants described *accountability* as the need to take responsibility for addressing unintended consequences and ensuring clear communication about system performance. This principle was associated with stakeholder expectations and crisis management.

Finally, we found no direct references to *privacy* and *security* in participants' narratives. These principles were not explicitly mentioned, suggesting they may not have been a significant focus in the specific practices discussed in the context of this study. Table I demonstrates evidence identified within the interviews.

TABLE I
QUOTATIONS ILLUSTRATING RESPONSIBLE AI PRINCIPLES IN SOFTWARE PROJECTS

| Principle | Evidence Examples |
| --- | --- |
| *Fairness* | "The bias, we tried to mitigate the bias of the image, of the representation of the person." (P07) |
| | "Why are we just making a video with people of the same style if I need the model to train various scenarios, several people?." (P10) |
| *Inclusiveness* | "They paid for us to have sign language classes ... to be able to emerge, think and talk to the interpreter." (P18) |
| | "I try to bring everyone that I want to work there ... bring all the situations that can happen so that it can have it there controlled in the laboratory." (P11) |
| *Reliability and Safety* | "The test structure ... as you are testing ... my machine works, then you take it on a day for a presentation, it does not work." (P06) |
| | "I think the process of building AI is continuous... From the feedback, we try to correct where the bias is." (P21) |
| *Transparency* | "In terms of testing, it's very complicated ... You don't know what the model will respond to results." (P08) |
| | "So today, for example, it has developed ... White box algorithms ... which are interpretable, they are explainable ... Facilitating the audit of the algorithms." (P18) |
| *Accountability* | "If it happens, it becomes much more of a crisis management situation, meaning making it very clear that the solution is in development." (P04) |

### B. Software Development Practices for Responsible AI

Our analysis demonstrated a series of practices that software professionals apply in their daily routine developing AI-powered systems, supporting compliance with *Responsible AI principles*. For instance, to address *fairness*, participants described practices aimed at mitigating bias in datasets and models, including *balancing datasets* and ensuring *diverse data representation* in model training. They also referred to techniques like *data augmentation* and *data weighting* to improve fairness.

Moreover, regarding *inclusiveness*, participants highlighted efforts to incorporate diverse perspectives during the development of solutions. This included *engaging underrepresented groups* during system design and testing, and *improving team diversity*. These practices aimed to simulate varied user scenarios and ensure accessibility, emphasizing the importance of designing systems that reflect the needs of a broad range of users.

Following this, practices to ensure *reliability and safety* were reported to involve *iterative testing* and *continuous improvement*. Participants referred to refining models based on user feedback and testing outcomes to ensure robustness and avoid misguided decisions across different contexts. Also, practices to enhance *transparency* focused on using *interpretable models and tools* that enable stakeholders to understand AI decisions. Finally, participants described practices linked to managing *accountability*, such as *engaging in clear communication with stakeholders* and *revisiting implemented models and solutions* to address unintended consequences. Once again, there were no explicit practices mentioned by participants in relation to *privacy and security*.

### C. Challenges to Incorporate Responsible AI Principles in Software Projects

Developing AI-powered systems that align with *Responsible AI principles* involves addressing several main challenges identified in our analysis:

- **Data-related challenges:** Bias in training data, imbalanced datasets, and concept drift complicate efforts to ensure fairness and reliability. These issues arise from unrepresentative datasets, overrepresentation of certain groups, and changing data distributions over time, requiring continuous monitoring and adaptation.
- **Organizational and team challenges:** The lack of team diversity and difficulties in multidisciplinary collaboration hinder inclusiveness. Homogeneous teams struggle to identify biases, while cross-functional teams often face misaligned goals and communication issues.
- **Bias detection and mitigation:** Subtle biases often remain undetected until they affect performance, requiring resource-intensive iterative testing and refinement. This challenge is critical for ensuring reliability and safety in AI systems.
- **Resource constraints:** Limited access to quality data, time, and technical expertise are pervasive challenges. High costs of acquiring representative datasets and implementing bias correction frequently conflict with project deadlines and budgets.
- **Cultural and behavioral complexity:** Designing inclusive systems necessitates addressing diverse user needs and behaviors, especially for marginalized groups. Understanding and integrating cultural and behavioral differences remain a significant challenge.

Some of these challenges are tied to specific principles, such as data-related issues being central to fairness, organizational and team challenges impacting inclusiveness, and difficulties in bias detection and mitigation affecting reliability and safety. However, resource constraints and cultural complexity emerge as cross-cutting challenges, influencing efforts across multiple principles. These interconnected challenges highlight the multifaceted nature of the difficulties faced by software professionals striving to achieve Responsible AI.

TABLE II
QUOTATIONS ILLUSTRATING CHALLENGES IN RESPONSIBLE AI

| Challenge | Evidence Example |
| --- | --- |
| Data-related challenges | "What you find is that what is there does not represent the real world. Then you have to add this new data." (P03) |
| Organizational and team challenges | "Teams that are very tech-savvy ... they have more difficulty looking at things in context ... were focused on mathematical aspects of the algorithm." (P20) |
| Bias detection and mitigation | "In fact, we never had this stage of trying to identify whether there were really biases or not ... It is understood in the very assessments of the model's behavior." (P03) |
| Resource constraints | "It is also very expensive to have access to quality data ... and this is a team concern." (P17) |
| Cultural and behavioral complexity | "The idea of bringing artificial intelligence was to give voice to these groups so that they had autonomy within this process within these cultural norms." (P12) |

## IV. DISCUSSIONS

The principles of Responsible AI [4], [5] provide an ethical framework for guiding AI development. However, our findings reveal a gap between these ideals and the practices and challenges reported by practitioners. Efforts to ensure fairness, such as balancing datasets and mitigating bias, were often limited by persistent issues like concept drift and the high cost of acquiring diverse training data. Similarly, while inclusiveness was promoted through team diversity and user engagement, these efforts were hindered by organizational barriers such as inadequate collaboration. These gaps suggest that current practices in the software industry often lack the resources or depth to fully address the challenges associated with operationalizing Responsible AI principles.

In this context, our study highlights the importance of aligning Responsible AI principles with realistic practices specific to software engineering, emphasizing the need for tailored tools and strategies to address foundational challenges such as resource constraints and diversity complexity in software projects. The identified cross-cutting challenges hinder progress in implementing multiple principles, underscoring the need for an integrated approach to bridge the gap between ethical guidelines and practical implementation. This calls for interdisciplinary collaboration and adaptable practices, ensuring that Responsible AI development in software engineering evolves alongside technological advancements and societal needs.

We acknowledge potential threats to validity in our study, particularly regarding generalization, framing, and methodological choices. As a qualitative study, our findings are not statistically generalizable but are instead transferable to similar contexts. While we ensured data saturation through in-depth interviews, the insights may not fully represent the breadth of industry practices. Additionally, the absence of direct questions about Responsible AI principles in the interviews may have limited the depth of responses on this topic, as participants addressed related concepts organically rather than explicitly. Finally, the use of GPT-4 Omni for qualitative analysis introduces potential risks of bias or inconsistency due to the generative nature of AI. We mitigated these risks through manual validation and accuracy checks.

## V. CONCLUSIONS

This study highlights the intricate relationship between Responsible AI principles and the practices and challenges faced by software engineering practitioners in AI development. Our findings reveal that while practitioners employ various strategies to align with these principles, persistent challenges demonstrate gaps in current industrial practices. These challenges call for tailored tools, interdisciplinary collaboration, and more inclusive frameworks to better support Responsible AI development. Future investigations should explore how to operationalize these principles across diverse software engineering contexts and investigate the role of organizational culture and policy in mitigating issues in the process.

## REFERENCES

[1] Z. Chen, M. Wu, A. Chan, X. Li, and Y.-S. Ong, "Survey on ai sustainability: emerging trends on learning algorithms and research challenges," *IEEE Computational Intelligence Magazine*, vol. 18, no. 2, pp. 60–77, 2023.

[2] M. K. Lee, N. Grgić-Hlača, M. C. Tschantz, R. Binns, A. Weller, M. Carney, and K. Inkpen, "Human-centered approaches to fair and responsible ai," in *Extended Abstracts of the 2020 CHI Conference on Human Factors in Computing Systems*, 2020, pp. 1–8.

[3] B. Brumen, S. Göllner, and M. Tropmann-Frick, "Aspects and views on responsible artificial intelligence," in *International Conference on Machine Learning, Optimization, and Data Science*. Springer, 2022, pp. 384–398.

[4] D. Schiff, B. Rakova, A. Ayesh, A. Fanti, and M. Lennon, "Principles to practices for responsible ai: closing the gap," *arXiv preprint arXiv:2006.04707*, 2020.

[5] A. Responsible, "principles from microsoft," 2022.

[6] Q. Lu, L. Zhu, X. Xu, J. Whittle, and Z. Xing, "Towards a roadmap on software engineering for responsible ai," in *Proceedings of the 1st International Conference on AI Engineering: software Engineering for AI*, 2022, pp. 101–112.

[7] N. Díaz-Rodríguez, J. Del Ser, M. Coeckelbergh, M. L. de Prado, E. Herrera-Viedma, and F. Herrera, "Connecting the dots in trustworthy artificial intelligence: From ai principles, ethics, and key requirements to responsible ai systems and regulation," *Information Fusion*, vol. 99, p. 101896, 2023.

[8] A. Deshpande and H. Sharp, "Responsible ai systems: who are the stakeholders?" in *Proceedings of the 2022 AAAI/ACM Conference on AI, Ethics, and Society*, 2022, pp. 227–236.

[9] Q. Wang, M. Madaio, S. Kane, S. Kapania, M. Terry, and L. Wilcox, "Designing responsible ai: Adaptations of ux practice to meet responsible ai challenges," in *Proceedings of the 2023 CHI Conference on Human Factors in Computing Systems*, 2023, pp. 1–16.

[10] Q. Lu, L. Zhu, J. Whittle, X. Xu *et al.*, *Responsible AI: Best Practices for Creating Trustworthy AI Systems*. Addison-Wesley Professional, 2023.

[11] Z. J. Wang, C. Kulkarni, L. Wilcox, M. Terry, and M. Madaio, "Farsight: Fostering responsible ai awareness during ai application prototyping," in *Proceedings of the CHI Conference on Human Factors in Computing Systems*, 2024, pp. 1–40.

[12] S. Baltes and P. Ralph, "Sampling in software engineering research: A critical review and guidelines," *Empirical Software Engineering*, vol. 27, no. 4, p. 94, 2022.

[13] K. Charmaz, *Constructing grounded theory*. sage, 2014.

[14] D. S. Cruzes and T. Dyba, "Recommended steps for thematic synthesis in software engineering," in *2011 international symposium on empirical software engineering and measurement*. IEEE, 2011, pp. 275–284.

[15] T. Heston and C. Khun, "Prompt engineering in medical education," *International Medical Education*, vol. 2, pp. 198–205, 8 2023.